\documentclass[aps,pra,showpacs,twocolumn,superscriptaddress, floatfix]{revtex4-1}
\usepackage{bm,color,amsmath,amssymb,mathrsfs,latexsym,graphicx,psfrag, epstopdf}
\usepackage[utf8]{inputenc}
\usepackage[normalem]{ulem}

\begin{document}
	\title{Phase transitions of the coherently coupled two-component Bose gas in a square optical lattice}
	\author{Ulrike Bornheimer}
	\affiliation{Centre for Quantum Technologies, National University of Singapore, 3 Science Drive 2, 117543 Singapore}
	\affiliation{MajuLab, CNRS-UNS-NUS-NTU International Joint Research Unit, UMI 3654, Singapore}
	\affiliation{Physics Department, Faculty of Science, National University of Singapore, 2 Science Drive 3, 117551 Singapore}
	\affiliation{Institut für Theoretische Physik, Goethe-Universität Frankfurt am Main, Max-von-Laue-Straße 1, 60438 Frankfurt am Main, Germany}
	
	\author{Ivana Vasi\'{c}}
	\affiliation{Scientific Computing Laboratory, Center for the Study of Complex Systems, Institute of Physics Belgrade, University of Belgrade, Pregrevica 118, 11080 Belgrade, Serbia}

	\author{Walter Hofstetter}
	\affiliation{Institut für Theoretische Physik, Goethe-Universität Frankfurt am Main, Max-von-Laue-Straße 1, 60438 Frankfurt am Main, Germany}

\begin{abstract}
We investigate properties of an ultracold, two-component bosonic gas in a square optical lattice at unit filling. In addition to density-density interactions, the atoms are subject to coherent light-matter interactions that couple different internal states. We examine the influence of this coherent coupling on the system and its quantum phases by using Gutzwiller mean field theory as well as bosonic dynamical mean field theory. We find that the interplay of strong inter-species repulsion and coherent coupling affects the Mott insulator to superfluid transition and shifts the tip of the Mott lobe toward higher values of the tunneling amplitude. In the strongly interacting Mott regime, the resulting Bose-Hubbard model can be mapped onto an effective spin Hamiltonian that offers additional insights into the observed phenomena.
\end{abstract}

\maketitle

\section{Introduction}

	Due to their highly controllable properties, systems of ultra-cold atoms are promising platforms for quantum simulations.
	One of the early successes in this direction was the observation of a superfluid-Mott insulator transition in a lattice Bose gas \cite{Greiner, Bloch2008}, as a prototype of a quantum phase transition. With recent advances in experimental techniques, present-day cold atom experiments feature finite-range interactions, for example in Rydberg dressed systems \cite{Loew, Schauss}, as well as artificial gauge potentials that mimic magnetic fields \cite{Dalibard, Goldman}. All these achievements bring these setups closer to simulating complex condensed matter systems.
	
	Multi-component systems, such as mixtures of different atoms or different hyperfine states of the same atomic species, introduce additional degrees of freedom that can be treated as pseudo spin.
	In the weakly interacting limit, depending on the ratio of intra-component and inter-component interactions, two-component mixtures may exhibit phase separation \cite{Pet2002}.
	The process of phase separation is substantially altered by introducing a coherent coupling term that enables a conversion of one internal atomic state into the other \cite{Blakie, Search, Tommasini, Lee, Merhasin, Abad2013, Butera, Chen, Shchedrin}.
	In advanced experiments this conversion is commonly implemented by an external laser that couples two internal atomic states  as, for example, in Rydberg systems \cite{Loew}. 
	 
	Effects of the coherent coupling are much less explored in the regime of strong repulsive interactions. In this paper, we address this topic in detail for the case of a two-component Bose gas on a square lattice. Strong interactions drive a transition from a condensate into a Mott insulator state, and in combination with coherent coupling, the system gives rise to a rich phase diagram that we investigate below.
	The subject has received some attention recently: DMRG studies for one-dimensional lattices have been performed \cite{Barbiero, Zhan2014} and the  possibility of an effective $xy$-antiferromagnetic state in two dimensions has been explored \cite{Grass2014}. In the recent papers \cite{ParnyPRL, Parny} it was shown that the coherent coupling between atoms and molecules can stabilize an additional quantum phase.
	
	The remainder of the paper is organized as follows: in the next section we introduce the main model of our study and briefly outline the methods that we use throughout the paper. In Sec.~III we address the case of finite bosonic coherences. 
	In particular we examine the neutral to polarized phase transition on top of the underlying condensate. The results for the case of stronger local interactions that lead to the superfluid-Mott transition with coherent coupling are presented in Sec.~IV. The case of imbalanced hopping amplitudes is the subject of Sec.~V. Finally, our main conclusions are summarized in Sec.~VI.

\section{Model and Methods}
	
	We investigate the phase diagram of an extended Hubbard model describing a two-component Bose gas in a square optical lattice with an additional  coherent coupling term. In second quantization the model explicitly reads
\begin{align}
	\hat{H}_{BH}=&-\sum_{<i,j>}^L\left( t_a\hat{a}^{\dagger}_i\hat{a}^{\phantom{\dagger}}_j+t_b\hat{b}^{\dagger}_i\hat{b}^{\phantom{\dagger}}_j+\mathrm{h.c.}\right)\nonumber\\
				&+\frac{1}{2}\sum_i^L\left( U_a\hat{n}_{ia}(\hat{n}_{ia}-1)+U_b\hat{n}_{ib}(\hat{n}_{ib}-1)\right)\nonumber\\		
				&+\sum_i^L\left(U_{ab}\hat{n}_{ia}\hat{n}_{ib}-\Omega\hat{a}^{\dagger}_i\hat{b}^{\phantom{\dagger}}_i+\mathrm{h.c.}\right),\label{eq:hamiltonian}
\end{align}
where $\hat{a}_i^{(\dagger)}$ and $\hat{b}_i^{(\dagger)}$ are the annihilation (creation) operators for bosonic species $a$ and $b$ respectively, and $\hat{n}_{ia}$ and $\hat{n}_{ib}$ are their number operators. The tight binding hopping amplitudes  of the respective species are denoted as $t_a$ and $t_b$ and $U_a$ and $U_b$ are their on-site interaction. Interactions are generally assumed to be local and repulsive in this study. An on-site interaction between particles of different species is designated as $U_{ab}$ and the term proportional to $\Omega$ allows for the conversion of one bosonic species into another on the same site. This last term is called the coherent coupling term. Local terms in the Hubbard Hamiltonian are summed over lattice sites $1, \ldots,L$, and in the first term  $<i,j>$ stands for the sum over all nearest neighbor sites $i$ and $j$ of the square lattice. Throughout the paper we will consider the case of $U_a = U_b = U$ and we set all scales by fixing $U=1$.
Our aim in this paper is to investigate possible ground states of the model (\ref{eq:hamiltonian}).

It is well known that by reducing the ratio of the tunneling amplitude $t_{a, b}$  over the repulsive interaction strength $U$ at commensurate lattice filling, bosons exhibit a superfluid to Mott-insulator transition \cite{Fisher1989, Bloch2008}. On top of this, an effective magnetic ordering emerges on the Mott side in two-component bosonic mixtures \cite{Altman2003, Kuklov2003}. These results are typically obtained for $\Omega = 0$ in the regime of   $\gamma \leq 1$, where we introduce the ratio between inter- and intra-species interaction $\gamma=U_{ab}/U$. In the opposite case of  $\gamma \geq 1$, it is energetically favorable for the two bosonic species to be spatially separated \cite{Pet2002}. However, a finite $\Omega$ will suppress this tendency by enabling a conversion between the two species. In this way, it allows us to address the regime $\gamma > 1$. Note that the model (\ref{eq:hamiltonian}) conserves only the total number of particles and not the  particle numbers of each species separately. In the limiting case $\gamma \gg 1$, only atoms of one species are present as the system avoids the high energy cost of $U_{ab}$.

This reasoning already suggests that in addition to the well understood superfluid to Mott phase transition, for the model (\ref{eq:hamiltonian}) we are able to distinguish another phase transition characterized by the polarization order parameter
\begin{equation}
\tilde{n}_i = \frac{\langle\hat{n}_{ia}\rangle-\langle\hat{n}_{ib}\rangle}{\langle\hat{n}_{ia}\rangle+\langle\hat{n}_{ib}\rangle}. 
\label{eq:polordpar}
\end{equation}
A strong interspecies interaction $U_{ab}$ favors a polarized phase with finite $\tilde{n}_i$. In contrast, the $\Omega$ term favors strong local coherence $\langle a_i^{\dagger} b_i +b_i^{\dagger}a_i\rangle$ that corresponds to a neutral phase with $\tilde{n}_i = 0$. To establish boundaries between different phases as a function of the physical parameters of the Hamiltonian (\ref{eq:hamiltonian}), we will use two approximate methods that we briefly outline here.

Features of the lattice Bose gas can be explored conveniently by means of the Gutzwiller mean-field theory \cite{Rokhsar1991, Krauth1992}, which amounts to decoupling non-local terms as
\begin{align}
	\hat{a}^{\dagger}_i\hat{a}^{\phantom{\dagger}}_j\approx\phi_i^*\hat{a}^{\phantom{\dagger}}_j+\phi_j\hat{a}^{\dagger}_i-\phi_i^*\phi_j,
\end{align}
where $\phi_i^{(*)}=\langle\hat{a}_i^{(\dagger)}\rangle$ is a condensate order parameter that is obtained in a self-consistent way. The approximation becomes an exact description in several limits: in the limit of infinite lattice coordination number $z$, in the atomic limit ($t = 0$) and in the weakly interacting limit, in the superfluid phase. However, for a vanishing condensate order parameter the lattice sites are completely decoupled whithin Gutzwiller mean-field theory and the description of the Mott domain is oversimplified. In order to go beyond this limitation, we will use bosonic dynamical mean field theory (BDMFT) \cite{Hubener2009f, Anders2010, Anders2011a, Li2011, Snoek2013}. 

Formally, BDMFT is derived as a second order expansion of the full model (\ref{eq:hamiltonian}) in terms of the inverse lattice coordination number. In comparison to Gutzwiller mean-field theory, we increase the order of the expansion by one \cite{Freericks2009}. The approximate effective problem obtained in this way is given by a bosonic Anderson impurity  model. Parameters of the effective model are set by imposing a self-consistency in terms of the condensate order parameter and the local Green's function. The main approximation is the assumption of the locality of self-energies, which is consistent with the second order expansion in the inverse of the coordination number \cite{Freericks2009,Metzner1989}. 
In order to take into account states that break translational symmetries, in this paper we use real-space BDMFT \cite{Li2011} and address a lattice size of $6 \times 6$, with a special focus on a possible two-sublattice ordering.

Both the Gutzwiller mean-field theory and BDMFT are implemented in the grand canonical ensemble. To this end, we introduce a single chemical potential $\mu$, $\hat{H}_{BH}\rightarrow\hat{H}_{BH}-\mu \sum_i^L \left(\hat{a}^{\dagger}_i\hat{a}^{\phantom{\dagger}}_i+\hat{b}^{\dagger}_i\hat{b}^{\phantom{\dagger}}_i \right)$, as the model (\ref{eq:hamiltonian}) conserves only the total number of particles. In the next sections we present and discuss results obtained by using these approaches for describing possible ground states of the model (\ref{eq:hamiltonian}).
\section{The Superfluid Regime}

\begin{figure*}
	\begin{center}
		\includegraphics[width = 0.93\textwidth]{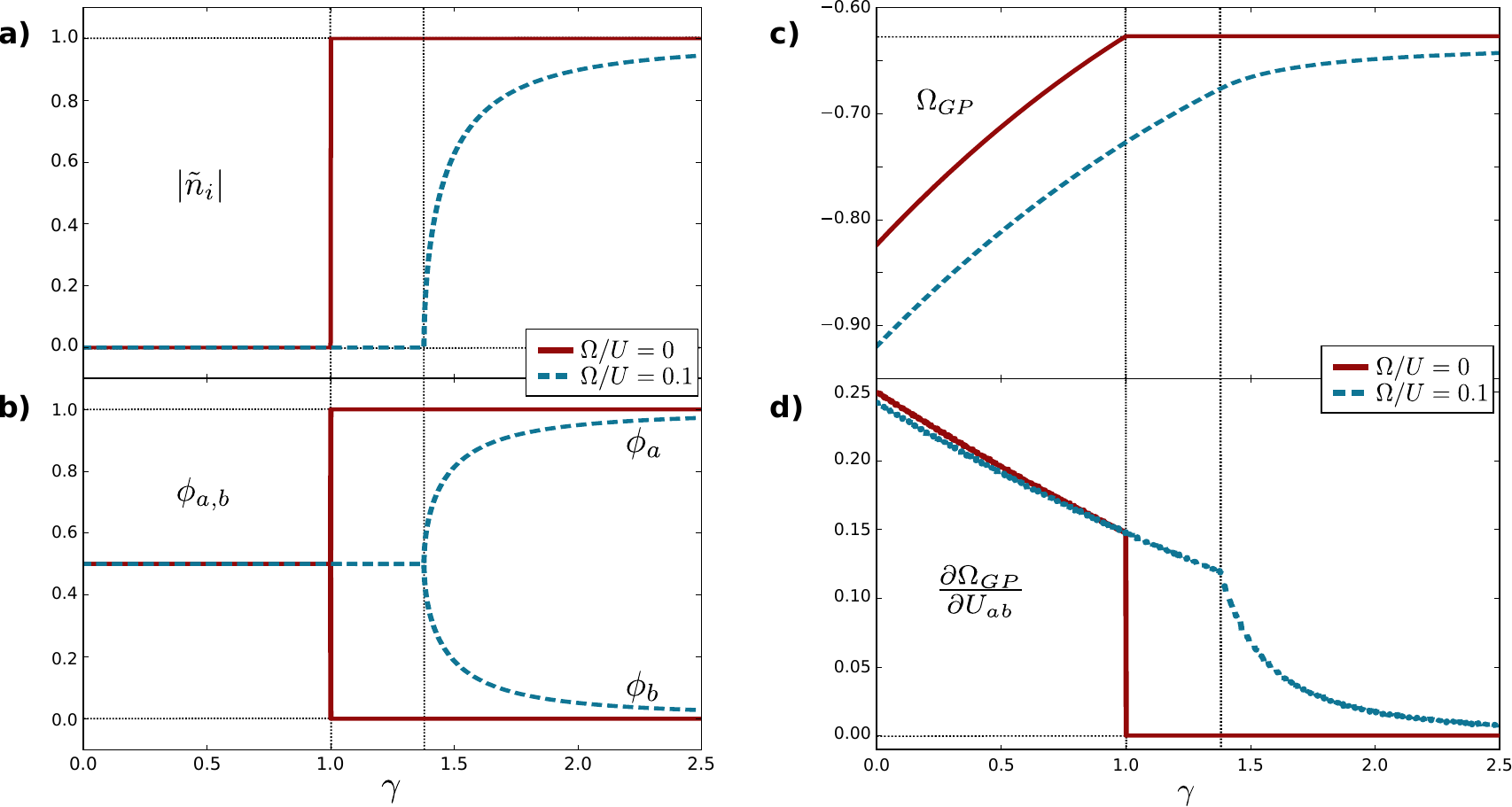}
    		\caption{a) Absolute value of the polarization order parameter $|\tilde{n}_i|$ defined in Eq.~(\ref{eq:polordpar}), b) condensate order parameters $\phi_a$ and $\phi_b$, c) lowest eigenvalue of the mean field Hamiltonian $\Omega_{GP}$, which corresponds to the grand canonical potential at zero temperature and d) its first derivative. These are plotted as a function of the interaction ratio $\gamma$ for unit  filling $n_a+n_b = 1$ and hopping amplitudes $t_{a} = t_{b} = 1/4$. \label{fig:abad}}
		\end{center}
		
	\end{figure*}

In this section we present the Gutzwiller analysis of the coherently coupled spinor Bose gas in the superfluid phase. In order to address states with a finite condensate fraction, we choose relatively high tunneling amplitudes $t_a/U = t_b/U = 1/4$. As the neutral to polarized phase transition is driven by the strong inter-component interactions $U_{ab}$,  we plot the polarization order parameter $\tilde{n}_i$, defined in Eq.~(\ref{eq:polordpar}), over the ratio $\gamma = U_{ab}/U$ in Fig.~\ref{fig:abad}a). 

Results for the model (\ref{eq:hamiltonian}) without coherent coupling are plotted for reference (red solid line) and the transition from $\tilde{n}_i = 0 $ to $|\tilde{n}_i| = 1$ is found at  $\gamma_c = 1$. We understand from previous works, that it is energetically favorable to have components of both species on each lattice site only in the case of weak inter-species repulsion. Strong interspecies repulsion leads to the polarized phase, where only particles of one species can be found on a single lattice site. For fixed atom densities the system will thus undergo phase separation. We notice that positive and negative values of $\langle\hat{n}_{a,i}\rangle-\langle\hat{n}_{b,i}\rangle$ ($\tilde{n}_i = \pm 1 $) appear  equally as results of numerical calculations with different initial conditions. This indicates two degenerate ground states in the polarized phase, as we will confirm in the following. In these calculations, initial parameters of the self-consistent loop or root search routines determine which of the ground states is selected, whereas in actual experiments the occurrence probabilities for both ground states are equal.\par	

\begin{figure}
    			\includegraphics[width = 0.9\linewidth]{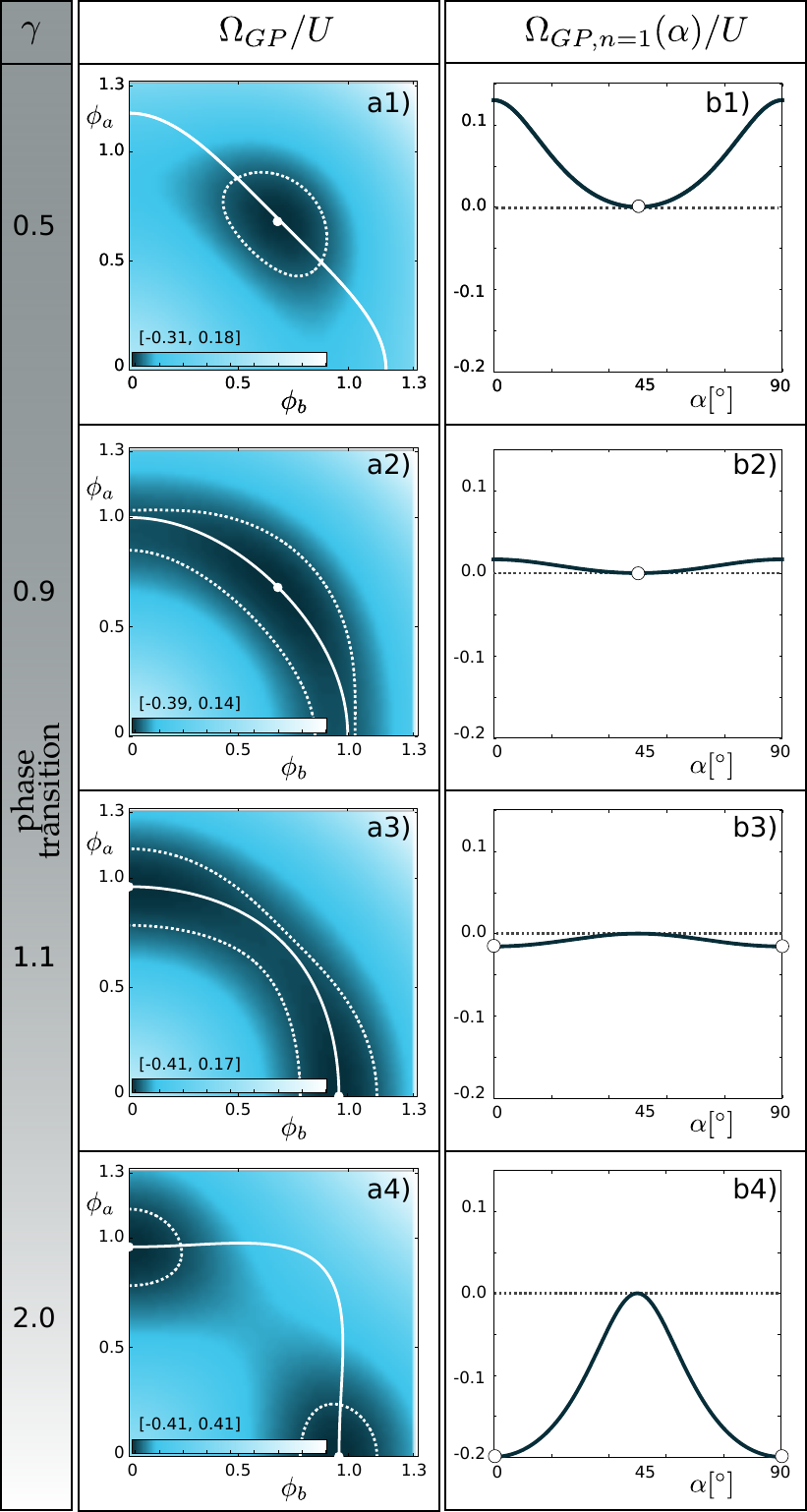}
    			\caption{\label{fig:phorder1} \textbf{Left series:} Lowest energy eigenvalue $\Omega_{GP}/U$ of the Gutzwiller mean-field Hamiltonian for $\Omega/U = 0$ plotted versus the condensate order parameters $\phi_a$ and $\phi_b$.  White data points mark the position of the minimum, white solid lines mark unit density and white dashed lines the area of $0.02$ around the minimum.  \textbf{Right series:} $\Omega_{GP}/U$ plotted along unit density lines (see white solid lines in left series) over the angle $\alpha=\arctan(\phi_a/\phi_b)$. Other Hamiltonian parameters are $t_a/U = t_b/U = 1/4$ and  $\mu/U = 1$. }
\end{figure}
We now turn to the effects of a finite coherent coupling  term and set $\Omega/U = 0.1$. While the polarization order parameter changes abruptly for vanishing coherent coupling (red solid line in Fig.~\ref{fig:abad}a), it exhibits a continuous change for finite coherent coupling (blue dashed line in Fig.~\ref{fig:abad}a).  Moreover, we notice that the coherent coupling shifts the transition point $\gamma_c$ to higher values of the inter-species interaction. 
The same qualitative behavior was reported in Refs.~\cite{Abad2013, Barbiero} in the  quasi one-dimensional geometry for the regime of weak interactions, both with and without a lattice. 
Within Gross-Pitaevskii theory, it was found analytically that the polarized phase sets in at $U_{ab}^c = U_{a,b} + 2\Omega/n$. 
For the parameters given in Fig.~\ref{fig:abad}, this relation yields a transition point at $\gamma_c = 1.2$. 
However, since the mentioned derivation is strictly valid only in the weakly interacting limit, where all bosons are condensed, the phase transition in Fig.~\ref{fig:abad} is expected to appear at a slightly  different value of $\gamma$.
In particular, we find that as the ratio $t/U$ is lowered further, the transition point between the polarized and neutral phase is shifted in favor of the neutral phase. The region of the neutral phase extends toward higher values of $\gamma $ and deviations with respect to the result obtained in the weakly interacting limit become more pronounced. This effect will be  further explored in the next section.
\par

The observation of discontinuities  in the polarization order parameter (Fig.~\ref{fig:abad}a)) draws our attention to the order of the observed phase transitions. We analyze this in Fig.~\ref{fig:abad}c) and d), where the grand potential and its first derivative are plotted as functions of $\gamma$. 
As we use the grand canonical description at zero temperature and explicitly include the chemical potential term, the grand potential is given by the expectation value of our mean-field Hamiltonian.
For $\Omega = 0$, we find that the first derivative of the grand potential is discontinuous at $\gamma_c$. This leads us to the conclusion, that the neutral to polarized phase transition is of first order for $\Omega = 0$.
In contrast, we observe a cusp in the same derivative for finite $\Omega$, implying that the phase transition is of second order.\par

\begin{figure}
    			\includegraphics[width = 0.9\linewidth]{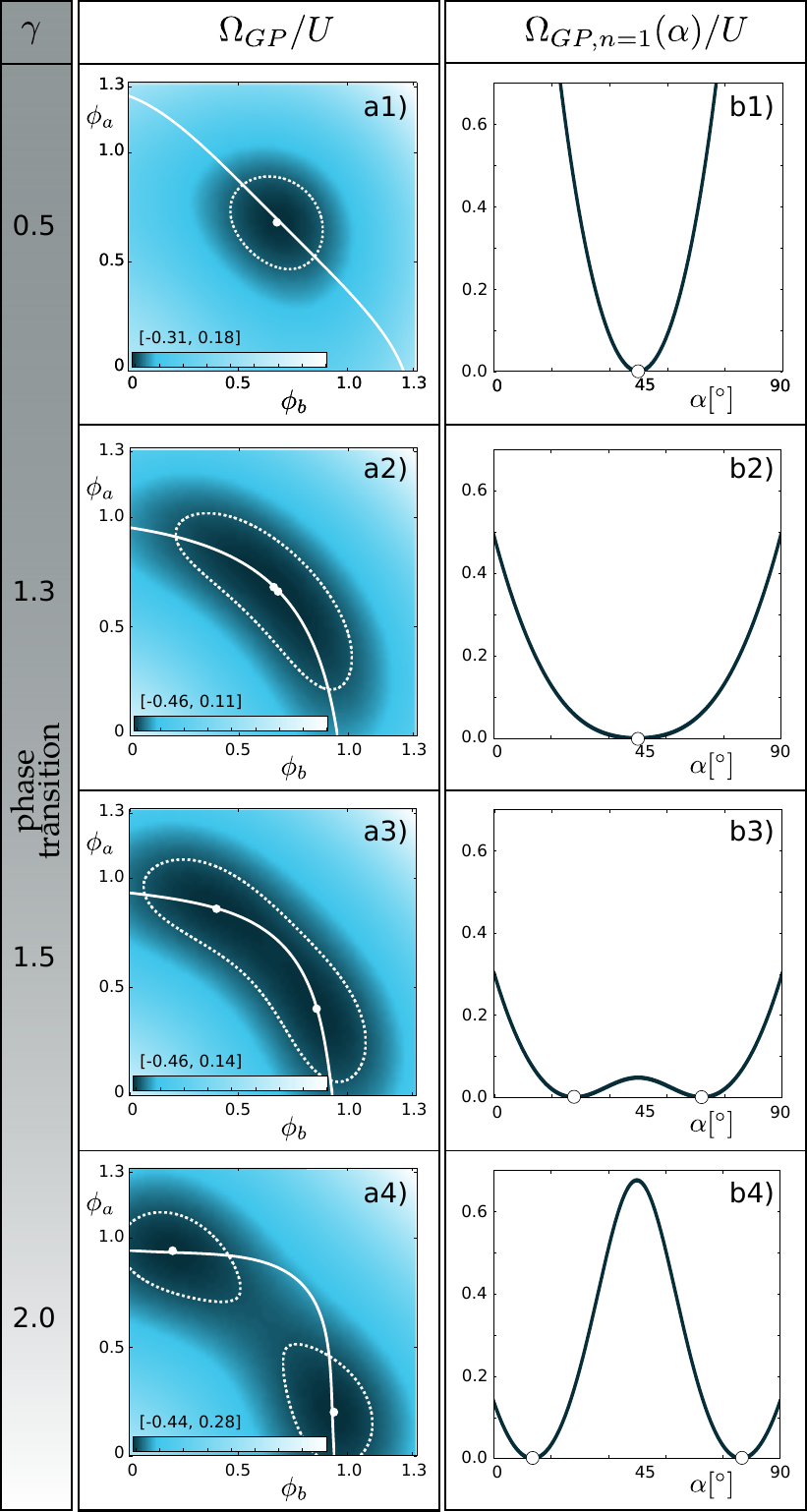}
    			\caption{\label{fig:phorder2}\textbf{Left series:} Lowest energy eigenvalue $\Omega_{GP}/U$ of the Gutzwiller mean-field Hamiltonian for $\Omega/U = 0.1$ plotted versus the condensate order parameters $\phi_a$ and $\phi_b$.  White data points mark the position of the minimum, white solid lines mark unit density and white dashed lines the area of $0.02$ around the minimum.  \textbf{Right series:} $\Omega_{GP}/U$ plotted along unit density lines (see white solid lines in left series) over the angle $\alpha=\arctan(\phi_a/\phi_b)$. Other Hamiltonian parameters are $t_a/U = t_b/U = 1/4$ and  $\mu/U = 1$. }
\end{figure}
To explore this in more detail, we plot the lowest energy eigenvalue of the mean field Hamiltonian as a function of the condensate order parameters in Figs.~\ref{fig:phorder1} and \ref{fig:phorder2} (left series). In both cases, for vanishing and for finite $\Omega$, we find a single energy minimum for the neutral phase at $\gamma< \gamma_c $, see Figs.~\ref{fig:phorder1}a1) and \ref{fig:phorder2}a1). The condensate order parameters $\phi_a$ and $\phi_b$ corresponding to this minimum are equal. In the polarized regime for $\gamma > \gamma_c$, however, two degenerate energy minima are present (Figs.~\ref{fig:phorder1}a4) and \ref{fig:phorder2}a4)). The degeneracy stems from the fact that the ground state breaks the symmetry between the two species, while the Hamiltonian is symmetric with respect to the interchange of these two species. In the polarized phase with $\Omega = 0$, the condensate order parameter of one of the species is strictly zero, while the other one has a finite value. In contrast, both order parameters are finite but non-equal at finite values of $\Omega$ as also shown in Fig.~\ref{fig:abad}b).

 	\begin{figure*}[!t]
\includegraphics[width=0.98\linewidth]{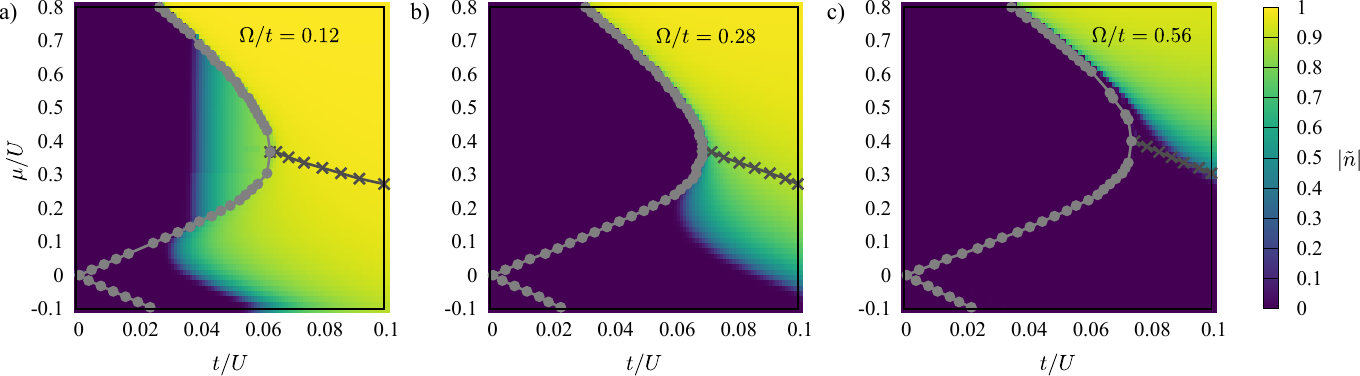}
\caption{Phase diagrams in the plane $t/U$-$\mu/U$ for $\gamma = U_{ab}/U = 8/5$. The color maps show the polarization $|\tilde n_i|$ defined in Eq. (\ref{eq:polordpar}) for a) $\Omega/t = 0.12$, b) $\Omega/t = 0.28$, and c) $\Omega/t = 0.56$. Dots show the Mott insulator to superfluid transition lines and crosses mark lines of constant density $\langle n_a+n_b \rangle=1$ on the superfluid side.}
\label{Fig:Figmut}
\end{figure*}
Having established the properties of the neutral and polarized phase, we now discuss the transition between them.\par

For $\Omega = 0$ and $\gamma = \gamma_c$, we find an infinitely degenerate energy minimum for constant $\sqrt{\phi_a^2+\phi_b^2}$. In the vicinity of the transition point, a single minimum at $\phi_a=\phi_b$ for $\gamma < \gamma_c$ abruptly transforms into two minima found at "distant" positions in the $\phi$-space for  $\gamma > \gamma_c$. The lowest eigenvalue of the Gutzwiller mean-field Hamiltonian plotted along the unit density line over $\alpha=\arctan(\phi_a/\phi_b)$ (right series of Fig.~\ref{fig:phorder1}) provides the reason for that. The energy has a parabolic shape around $\alpha=45^{\circ}$. The coefficient of this parabola changes from being positive in the neutral phase, which results into a minimum at $\alpha = 45^{\circ}$, to being negative in the polarized phase, which results into two well separated minima at $\alpha = 0^{\circ},90^{\circ}$.\par

At finite $\Omega$, we do not find an abrupt change in the minima at the transition point. The neutral minimum at $\phi_a = \phi_b$ splits into two degenerate minima that evolve towards their final values with increasing $\gamma$. Accordingly, the energy plot along the unity density line (right series, Fig.~\ref{fig:phorder2}) is not a parabola that simply flips the sign of its coefficient, but one that develops a bump at $\alpha = 45^{\circ}$ and thus gradually shifts its minima away from that point.\par


\section{Superfluid-Mott Transition}

Strong on-site interactions suppress density fluctuations and deplete the condensate. At commensurate densities, this mechanism drives a transition from a superfluid into a Mott insulator state.
In this section we map out the phase diagram of the model defined in Eq.~(\ref{eq:hamiltonian}) as a function of the tunneling amplitude $t/U$ and coherent coupling  $\Omega/t$ by using BDMFT  at zero temperature.

In the limit of vanishing coherent coupling $\Omega \rightarrow 0 $, for $\gamma > 1$ (inter-species interaction stronger than intra-species interaction) our BDMFT simulations, implemented in the grand-canonical ensemble, recover the well-known results for the Mott-insulator to superfluid transition for a single bosonic species on a square lattice \cite{Capogrosso2008, Anders2010}. At finite $\Omega$ and $t=0$ (the atomic limit)  and at a total filling $\langle n_{ai}+n_{bi}\rangle = 1$, the ground state has no polarization. This can easily be seen by considering a single-site Hamiltonian in the subspace spanned by $|1,0\rangle =a_i^{\dagger}|0\rangle$ and $|0,1\rangle =b_i^{\dagger}|0\rangle $, which is given by 
\begin{equation}
\left(
 \begin{array}{cc}
  -\mu & -\Omega\\
  -\Omega& -\mu
 \end{array}
\right).
\end{equation}
The ground state is $|\text{GS}\rangle = (|0,1\rangle+|1,0\rangle)/\sqrt{2} $ and it is neutral, since $\langle \text{GS}| n_{ai}-n_{bi}|\text{GS}\rangle=0$. In the following we investigate a range of tunneling amplitudes $t_a/U = t_b/U = t/U \in [0.001, 0.1]$ and coherent couplings $\Omega/t \in [0,1] $. We scan phase diagrams spanned by $t/U$ and $\mu/U$ at fixed $\Omega/t$ to access points with total unit filling as shown in Fig.~\ref{Fig:Figmut} for $\gamma=8/5$.

We begin our analysis  with a small fixed ratio $\Omega/t$. In Fig.~\ref{Fig:Figmut}a) we present the absolute value of  the polarization order parameter $|\tilde{n}|$  (Eq.~(\ref{eq:polordpar}))  for $\Omega/t = 0.12$.  Increasing the tunneling $t/U$ from a starting point near the atomic limit, we encounter a transition from the neutral into the polarized state, before reaching the tip of the first Mott lobe. With further increase of $t/U$, we find  a second transition from the polarized Mott state into the polarized condensate state (see Fig.~\ref{Fig:Figmut}a)). At stronger $\Omega/t \geq 0.28$, we find that the whole Mott lobe is neutral and around the tip of the lobe, we have a transition from the neutral Mott directly into the polarized superfluid. An example of this behavior is shown in  Fig.~\ref{Fig:Figmut}b). Finally, for $\Omega/t > 0.44$, there is a transition from the neutral Mott insulator into the neutral superfluid, followed by a second transition from a neutral into a polarized superfluid state, see Fig.~\ref{Fig:Figmut}c. The change in the polarization on the superfluid side of the diagrams as a function of the chemical potential $\mu$  (Figs.~\ref{Fig:Figmut}a-c, a vertical cut) can be understood as follows: by increasing the chemical potential $\mu$, the total density increases (not explicitly shown in figures) and the enhanced contribution of repulsive interactions can overcome the effect of the coupling $\Omega$ that favors a neutral state.

\begin{figure}
\includegraphics[width=0.98\linewidth]{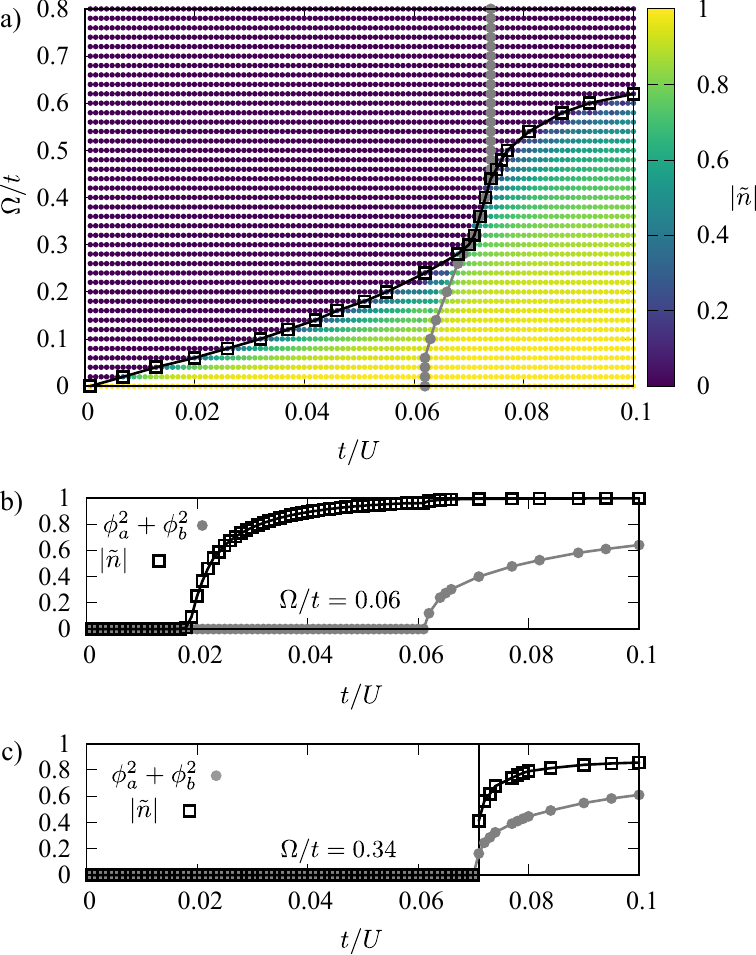}
\caption{a) The phase diagram of the model at unit filling for $\gamma = U_{ab}/U = 8/5$. The color map shows the polarization $|\tilde n|$ defined in Eq. (\ref{eq:polordpar}). Dots show the superfluid to Mott insulator transition line, and the squares give neutral to polarized transition line. Bottom plots: cuts through the phase diagram for b) $\Omega/t = 0.06$ and c) $\Omega/t = 0.34$. Condensate order parameters and polarization $|\tilde n|$ are plotted as functions of $t/U$ in b) and c).}
\label{Fig:Figpd}
\end{figure}
A complete phase diagram as a function of $\Omega/t$ and $t/U$ at unit filling obtained from the previous type of calculation is presented in Fig.~\ref{Fig:Figpd}a). The color plot gives the polarization $|\tilde n|$, the  black squares form a transition line between the polarized and unpolarized states, and  gray circles show the Mott insulator (left part) to superfluid (right part) transition line. The two transition lines, marking the Mott insulator to superfluid transition and, neutral to polarized transition, coincide in the intermediate region.
The Mott region extends up to $t/U \approx 0.06$ for a fully polarized Mott to polarized superfluid transition (weak $\Omega/t$), which is very close to the result for the Mott-lobe tip in a single-component case. At strong $\Omega/t$ we have a neutral Mott to neutral superfluid transition that takes place at a higher value of $t/U \approx 0.074$, as strong $U_{ab}$ plays a more pronounced role in this case. 
In the horizontal cut shown in Fig.~\ref{Fig:Figpd}b) we explicitly show that at weak $\Omega/t$ there are two second-order phase transitions. These transitions merge into a single transition point with a jump in polarization $|\tilde n| $ at intermediate $\Omega/t$, as shown in Fig.~\ref{Fig:Figpd}c). At even stronger $\Omega/t$, we find two separate transitions of the second order again.

In order to explain the neutral-polarized transition on the Mott side, we complement numerical BDMFT results with an insight obtained from an effective spin Hamiltonian, which is valid in the limit of strong interactions. The spin model is derived  via second order perturbation theory in the hopping amplitude and for unit filling. Starting from  model (\ref{eq:hamiltonian}),  it is expressed in the pseudo-spin basis $|\uparrow\rangle=|n_a=1,n_b=0\rangle$ and $|\downarrow\rangle=|n_a=0,n_b=1\rangle $ \cite{Kuklov2003, Altman2003, Petrescu2013}. Equivalently, one can also use the Schrieffer-Wolff transformation \cite{Schrieffer1966} to obtain the same effective Hamiltonian:
\begin{align}
	\hat{H}_{\text{eff}} = &-J_{zz}\sum_{<i,j>}\hat{S}^z_i\hat{S}^z_j-J_{\perp}\sum_{<i,j>}\left(\hat{S}^x_i\hat{S}^x_j + \hat{S}^y_i\hat{S}^y_j\right)\nonumber\\
	&-J_z\sum_i\hat{S}_i^z-2\Omega\sum_i\hat{S}^x_i,
	\label{eq:spinmodel5}
\end{align}
where we introduce $\hat{S}_i^{l} = (1/2)(\hat{a}_i^{\dagger}, \hat{b}_i^{\dagger}) \sigma^l \left(\begin{array}{c}\hat{a}_i\\ \hat{b}_i\end{array}\right)$ using the Pauli matrices $\sigma^l$ with $l = x,y, z$.
For the parameters considered in this section ($t_a/U=t_b/U=t/U$), the spin coupling constants simplify to
	$J_{zz} = 4t^2\left(2\gamma-1\right)/\gamma U$,
	$J_{\perp}=4t^2/\gamma U$ and
	$J_z =0.$
Based on the model (\ref{eq:spinmodel5}), in the strongly interacting limit of the Bose-Hubbard model with coherent coupling from Eq.~(\ref{eq:hamiltonian}), we expect to find different phases depending on the magnitude of the coefficients $J_{z}$, $J_{zz}$, $J_{\perp}$ and $\Omega$. In particular, the spin ordering along $z$-direction is equivalent to the finite polarization order parameter from Eq.~(\ref{eq:polordpar}), while spin alignment along $x$-direction corresponds to the neutral phase.
At the transition line of the neutral and polarized phase we expect the spin couplings $J_{\perp}$ and $\Omega$, that favor spin alignment in the $x$ direction, to be comparable to the $z$-ordering term $J_{zz}$. This reasoning leads to an approximate condition for the transition line
\begin{equation}
\frac{\Omega_c}{t} \propto \left(1-\frac{1}{\gamma}\right)\frac{t}{U}.
\label{eq:fit}
\end{equation}
We fit the numerical data according to this argumentation to $\Omega_c/t \propto \alpha t/U$,
where $\alpha$ is the fitting parameter \cite{Zhan2014}. For small $t/U$, we find $\Omega_c/t\approx 3.2 t/U$  for  $\gamma = 8/5$ (see Fig. \ref{Fig:Figpd}) and $\Omega_c/t\approx 2.15t/U$ for  $\gamma = 4/3$.  These fitting constants explicitly fulfill   the $(1-1/\gamma)$ dependence in Eq.~(\ref{eq:fit}).

\section{Imbalanced hopping amplitudes}

Up to now we considered two fully equivalent bosonic components described by $t_a = t_b$ and $U_a = U_b$. As a consequence,  two degenerate solutions with $\pm \tilde n$ were found in the polarized regime. In this section we address a more general case of imbalanced hopping amplitudes $t_a \neq t_b$. In particular, we investigate how this imbalance affects the neutral to polarized transition within the Mott regime at unit filling. We consider both positive and negative tunneling amplitudes. The latter case is less common, but it is experimentally accessible in shaken optical lattices \cite{Eckardt2010}.

\begin{figure}[!t]

\includegraphics[width = 0.5\textwidth]{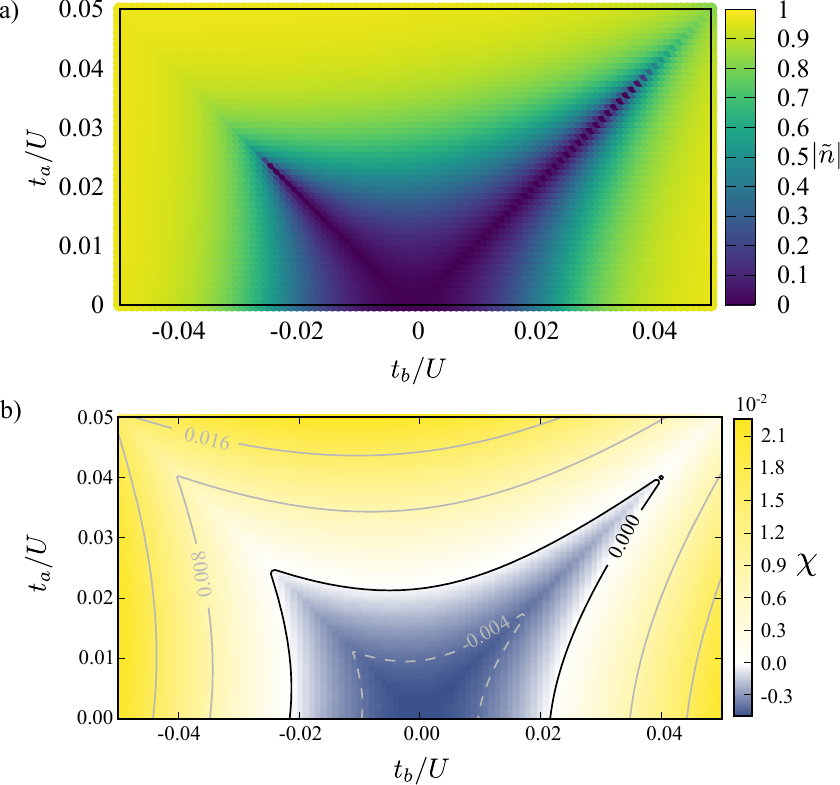}
	\caption{ a) Absolute value of the polarization order parameter $|\tilde n_i|$ in the $t_a-t_b$ plane obtained by BDMFT. b) Auxiliary function $\chi$ defined in Eq.~(\ref{eq:chi}). Parameters: $\gamma = U_{ab}/U = 8/5$, $\mu = 0.2 U_{ab}$, $\Omega/U = 5\times 10^{-3}$.\label{fig:pd_JaJb}}
\end{figure}
In Fig.~\ref{fig:pd_JaJb}a) we present the absolute value of the polarization order parameter $|\tilde n|$ as a function of tunneling rates $t_a$ and $t_b$. Other parameters of the Hamiltonian (\ref{eq:hamiltonian}) are set to $\gamma = U_{ab}/U = 8/5$, $\Omega/U = 5 \times 10^{-3}$ and $U_a = U_b = U$. We limit ourselves to small absolute values of the ratios  $|t_a|/U$ and $|t_b|/U$ and unit filling $\langle n_a + n_b\rangle =1 $, so that the system is in the Mott phase.  Low values of the polarization are found for very weak tunneling amplitudes, in the region given by $|t_{a, b}| < t_c$, where we have $t_c/U\approx 2.5 \times 10^{-3}$ for the data shown in Fig.~\ref{fig:pd_JaJb}. Along the diagonals $t_a = \pm t_b$, the neutral phase with $\tilde n = 0$ extends up to the largest values of $t_a$ and $t_b$. We notice a clear difference in the extension of the neutral phase for the case of $t_a = t_b$ in comparison with the case of $t_a = -t_b$.

In order to explain the features observed in Fig.~\ref{fig:pd_JaJb}a), we use the effective spin model defined in Eq.~(\ref{eq:spinmodel5}), that captures low energy properties of the Hamiltonian (\ref{eq:hamiltonian}) deep in the Mott domain.
For the parameters considered in this section, the coefficients of the spin model (\ref{eq:spinmodel5}) simplify to
\begin{align}
	J_{zz} &= 2\frac{t_a^2+t_b^2}{U}\left(2-\frac{1}{\gamma}\right),\label{eq:Jzz1}\\
	J_z &= 8\frac{t_a^2-t_b^2}{U},\label{eq:Jzz2}\\
	J_{\perp} &= \frac{4t_at_b}{\gamma U}. \label{eq:Jperp}
\end{align}
where we typically consider $1 < \gamma < 2$. According to Eq.~(\ref{eq:spinmodel5}), the coefficient $J_z$ plays the role of an effective chemical potential that selects which of the two species is preferred in the polarized phase.
We notice that $J_{zz}$ increases quadratically both with $t_a$ and $t_b$, which leads to the largest increment in this spin coupling term isotropically around $t_a=t_b=0$. In contrast, $J_z$ depends on the difference of the square of both hopping amplitudes and exhibits the strongest increase perpendicular to the diagonals $t_a=\pm t_b$. 
In our analysis, the value of $\Omega$ is kept constant. Thus, it is the dominant quantity in the region around $t_{a,b}\approx0$, where all other spin couplings (\ref{eq:Jzz1})-(\ref{eq:Jperp}) are weak and where we find the neutral phase accordingly. The asymmetry between the negative and positive side of $t_a$ in the plot shown in Fig.~\ref{fig:pd_JaJb}a) arises due to the $J_{\perp}$ coupling.
For  hopping amplitudes of the same sign, the $J_{\perp}$ coupling is positive. As such, it lowers the energy of the neutral phase and thus shifts the phase transition to higher values of $t_{a,b}$. The opposite is true for hopping amplitudes of different sign.

To sum up the implications of the spin model and compare these to our numerical results, we examine the interplay of spin coupling amplitudes with the help of an auxiliary function $\chi$. All spin couplings that favor the polarized phase are marked with a positive sign, while the spin couplings favoring the neutral phase carry a negative sign
\begin{align}
	\chi = J_{zz}+|J_z|-J_{\perp}-\Omega.\label{eq:chi}
\end{align}
The roots of this function give an estimate for the neutral to polarized phase transition line and the resulting plot in Fig. \ref{fig:pd_JaJb}b) qualitatively recovers the structure of the numerical phase diagram.

\begin{figure}[t!]
	\begin{center}
                \includegraphics[width = 0.5\textwidth]{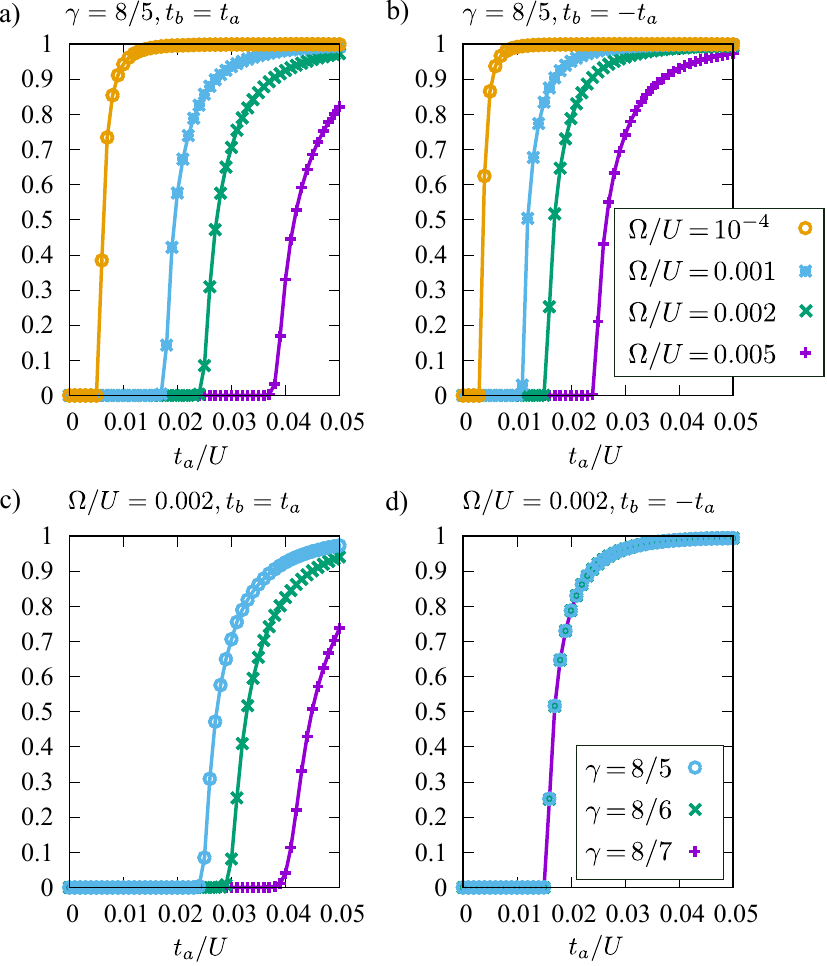}
		\caption{
		Absolute value of the polarization order parameter $|\tilde n_i|$ as a function of the hopping amplitude for $t_b = t_a$ (left column) and $t_b = -t_a$ (right column). Parameters not shown in the plot: $U = 1, \mu = 0.2 U_{ab}$. In the two top plots a) and b), the value of the coherent coupling term is varied. 
		In the two bottom plots, c) and d), we vary the value of $\gamma$. For $t_a = -t_b $ we find that our results do not depend on the value of $\gamma$. \label{fig:Fig7}
		}
	\end{center}
\end{figure}

We now investigate how the transitions from Fig.~{\ref{fig:pd_JaJb}} are affected by the change in the interaction ratio $\gamma$ and in the coherent coupling $\Omega$. It turns out that the corresponding $t_a-t_b$ plots look qualitatively similar to the plot  Fig.~{\ref{fig:pd_JaJb}}a). In order to make a quantitative comparison,  we plot the absolute value of the polarization $|\tilde n|$ as a function of $t_a$ for $t_a = t_b$ (Fig.~\ref{fig:Fig7}, left column) and for $t_b = -t_a$ (Fig.~\ref{fig:Fig7}, right column). In the plots shown in Fig.~\ref{fig:Fig7}a) and b), we set $\gamma = 8/5$ and vary $\Omega$. Our results show good agreement  with the expectation $t_c/U \sim \sqrt{\Omega/U}$ from Eq.~(\ref{eq:fit}) for the transition point $t_c$. For the two cases considered ($t_a = t_b$ and $t_a = -t_b$) only the proportionality constants are different.

In the plots presented in Fig.~\ref{fig:Fig7}c) and d) we keep the value  $\Omega/U = 2 \times 10^{-3} $ fixed and change $\gamma$. For $t_a = t_b $ our results are well approximated by $t_c \sim \left(1-1/\gamma\right)^{-1/2}$, in agreement with Eq.~(\ref{eq:fit}). 
In contrast, for $t_a = -t_b $ we find that our results do not depend on the value of $\gamma$. 
This can be explained by looking at the spin coupling constants from Eqs.~(\ref{eq:Jzz1})-(\ref{eq:chi}). As mentioned earlier,  the $J_{\perp}$ coupling opposes the impact of the $\Omega$ term for hopping amplitudes of different sign. The relevant contribution to the auxiliary function $\chi$ (defined in Eq.~(\ref{eq:chi})) that leads to the aforementioned $\gamma$ invariance, however, is given by the sum $J_{zz}+ |J_{\perp}|$. The $\gamma$ parts of both terms cancel, leaving $\chi$ independent of $\gamma$.

\section{Summary and Discussion}

In this work, we focused on a two-component mixture of coherently coupled bosons in a square optical lattice. We analyzed the phase transition between the polarized (finite $ \tilde{n}_i$, Eq.~(\ref{eq:polordpar})) and the neutral phase, driven by an interplay of the coherent coupling $\Omega$ and the interspecies repulsion $U_{ab}$ at unit filling $\langle n_{ia} + n_{ib} \rangle = 1$. 

By comparing Gutzwiller results with more demanding BDMFT calculations, we found that the former provide a reasonable description of the system in the superfluid regime. We investigated the energy landscape of the mean-field Hamiltonian as a function of the two condensate order parameters and established that the coherent coupling leads to a second order phase transition between the polarized and the neutral phase.  
Furthermore, we found that the neutral phase is suppressed as the ratio of inter- and intra-species interactions $\gamma$ increases.\par

On the Mott side of the phase diagram, where BDMFT calculations provide a necessary extension of the simpler mean-field theory, we found the polarized phase to be favored only at very low values of the coherent coupling (Fig.~\ref{Fig:Figpd}a). 
From this, we concluded that the long-range order of the condensate seems to favour the polarized phase. 
To better understand our numerical results in the Mott phase, we derived an effective spin model using second order perturbation theory in the hopping amplitudes. From the coupling constants of the effective model we inferred that the polarized to neutral transition line is approximately given by $\frac{\Omega^c}{t} \propto \left(1-1/\gamma\right)\frac{t}{U}$ in very good agreement with our numerical results (Fig. \ref{Fig:Figpd}). The dominance of the neutral phase in the deep Mott regime at unit filling agrees well with the other findings in this paper, as well as in the literature, especially with recent DMRG calculations \cite{Zhan2014}. Furthermore, our BDMFT results indicate three possible transitions with increasing $t/U$,  depending on the value of coherent coupling $\Omega$ and interaction ratio $\gamma = U_{ab}/U$: polarized Mott states turn into a polarized superfluid, the polarized Mott phase turns directly into a neutral superfluid, and from the neutral Mott phase there is a transition to a neutral superfluid.  The tip of the Mott lobe is positioned at the smallest value of $t/U$ in the first case, while the lobe extends up to the largest value of the tunneling amplitude for the third case.\par

Finally, we explored the effects of imbalanced hopping amplitudes for the two species for both (common) positive and negative tunneling terms. The latter were  realized only recently in periodically driven optical lattices \cite{Eckardt2010}. 
For now, we considered the case of $\gamma > 1$ and found that $\Omega$, enforcing the neutral phase, has a strong influence only for very small hopping amplitudes or for $t_a=\pm t_{b}$. An interesting asymmetry, that shows up in the two latter cases, was traced back to the sign change of one of the coupling constants in the effective spin model. In future work, we plan to consider the  case of $\gamma=1/2$ and $t_a = -t_b$, where recent calculations \cite{Grass2014} suggested an occurrence of the $xy$-antiferromagnetic phase.
	
\section{Acknowledgments}

Support by the Deutsche Forschungsgemeinschaft via DFG SFB/TR 49, DFG FOR 801, DFG SPP 1929 GiRyd and the high-performance computing center LOEWE-CSC is gratefully acknowledged.  
This work was supported in part by DAAD (German Academic and Exchange Service) under project BKMH.
I.~V.~acknowledges support by the Ministry of Education, Science, and Technological Development of the Republic of Serbia under projects ON171017 and BKMH, and by the European Commission under H2020 project VI-SEEM, Grant No. 675121. Numerical simulations were partly run on the PARADOX supercomputing facility at the Scientific Computing Laboratory of the Institute of Physics Belgrade.
U.~B.~acknowledges support by the CQT PhD program. The Centre for Quantum Technologies (CQT) is a Research Centre of Excellence founded by the Ministry
of Education, Singapore and the Singapore National Research Foundation.

%

\end{document}